\documentclass[a4paper,UKenglish,cleveref, autoref, thm-restate]{lipics-v2021}

\pdfoutput=1 %
\hideLIPIcs  %

\def\MdR{\ensuremath{\mathbb{R}}}

\newcommand{\ouralg}{\textsc{BuffCut}\xspace}
\definecolor{shade1}{gray}{0.95} %
\definecolor{shade2}{gray}{0.99} %
\definecolor{best}{RGB}{220,235,255}

\usepackage{amsbsy}
\usepackage{mathtools}
\usepackage{xspace}
\usepackage{graphicx}
\usepackage{subcaption}
\usepackage{booktabs}
\usepackage{multirow}
\usepackage[table]{xcolor} 
\usepackage{multirow}
\usepackage{tikz}
\usepackage[most]{tcolorbox}
\usepackage[ruled,vlined]{algorithm2e}

\newtcolorbox{takeaway}{
	colback=gray!6,
	colframe=black!55,
	boxrule=0.6pt,
	arc=2pt,
	left=6pt,
	right=6pt,
	top=5pt,
	bottom=5pt,
}

\title{BuffCut: Prioritized Buffered Streaming Graph Partitioning} %

\author{Linus {Baumgärtner}}{Heidelberg University, Germany}{lbaumgaertner@culba.de}{https://orcid.org/0009-0002-1488-4663}{}
\author{Adil {Chhabra}}{Heidelberg University, Germany }{adil.chhabra@informatik.uni-heidelberg.de}{https://orcid.org/0009-0009-5726-9389}{}
\author{Marcelo Fonseca {Faraj}}{A+W Software, Germany }{marcelofaraj@gmail.com}{https://orcid.org/0000-0001-7100-236X}{}
\author{Christian {Schulz}}{Heidelberg University, Germany}{christian.schulz@informatik.uni-heidelberg.de}{https://orcid.org/0000-0002-2823-3506}{}

\authorrunning{L. Baumgärtner et al.} %

\Copyright{Linus {Baumgärtner}, Adil {Chhabra}, Marcelo Fonseca {Faraj} and Christian {Schulz}} %



\category{} %

\relatedversion{} %

\supplement{Source Code: \url{https://github.com/KaHIP/HeiStream}}%

\acknowledgements{We acknowledge support by DFG grant SCHU 2567/5-1. 
	Moreover, we would like to acknowledge Dagstuhl Seminar 23331 on Recent Trends in Graph Decomposition.
}

\nolinenumbers %

\EventEditors{John Q. Open and Joan R. Access}
\EventNoEds{2}
\EventLongTitle{42nd Conference on Very Important Topics (SEA 2016)}
\EventShortTitle{SEA 2026}
\EventAcronym{SEA}
\EventYear{2026}
\EventDate{December 24--27, 2026}
\EventLocation{Little Whinging, United Kingdom}
\EventLogo{}
\SeriesVolume{42}
\ArticleNo{23}

\begin{document}
	
	\maketitle
	
	\begin{abstract} Streaming graph partitioners enable resource-efficient and massively scalable partitioning, but one-pass assignment heuristics are highly sensitive to stream order and often yield substantially higher edge cuts than in-memory methods. 
		We present \ouralg, a buffered streaming partitioner that narrows this quality gap, particularly when stream ordering is adversarial, by combining prioritized buffering with batch-wise multilevel assignment. 
		\ouralg maintains a bounded priority buffer to delay poorly informed decisions and regulate the order in which nodes are considered for assignment. 
		It incrementally constructs high-locality batches of configurable size by iteratively inserting the highest-priority nodes from the buffer into the batch, effectively recovering locality structure from the stream. 
		Each batch is then assigned via a multilevel partitioning algorithm. 
		Experiments on diverse real-world and synthetic graphs show that \ouralg consistently outperforms state-of-the-art buffered streaming methods. Compared to the strongest prioritized buffering baseline, \ouralg achieves 20.8\% fewer edge cuts while running 2.9$\times$ faster and using 11.3$\times$ less memory. Against the next-best buffered method, it reduces edge cut by 15.8\% with only modest overheads of 1.8$\times$ runtime and 1.09$\times$ memory. 
	\end{abstract}

	\section{Introduction}
	\label{sec:introduction}
	Graphs are a central abstraction for modeling complex relationships in modern data-driven systems. Massive graphs with billions of nodes and edges are routinely used to represent social, biological, navigational, and technical networks, enabling tasks such as community detection, pathway analysis, and recommendation~\cite{alpert1995rdn,alpert1999spectral,catalyuerek1996dis,DellingGPW11,heuvelinecoop,george1973nested,Lau04}. 
	More recently, graphs have become a core data structure in emerging applications such as real-time analytics over social networks and financial transaction streams, and machine learning models based on graph neural networks~\cite{timeevolvinggraphprocessing,gnnfrauddetection}. These graphs see large numbers of edges and nodes added or removed frequently and have highly heterogeneous degree distributions, placing pressure on both memory and computation. Thus, graph processing systems must balance a trade-off between delivering high-quality analytics while operating under tight memory and runtime constraints, often in online or near-real-time settings.
	
	To address these challenges, graph processing is typically performed in distributed environments. The input graph is divided into disjoint blocks that are stored and processed across multiple workers, which communicate along edges that cross block boundaries. In this setting, inter-block edges dominate communication and synchronization costs, while imbalanced partitions create stragglers that slow parallel execution. Graph partitioning is therefore a fundamental step in distributed graph processing: it aims to divide a graph into roughly equal-sized subgraphs while minimizing inter-block connectivity. However, graph partitioning is NP-hard~\cite{bourse-2014,Garey1974}, and no constant-factor approximation exists for general graphs unless $P=NP$~\cite{BuiJ92}, necessitating the use of heuristic algorithms in practice.
	
	State-of-the-art in-memory partitioners such as \textsc{METIS}~\cite{karypis1998fast}, \textsc{KaHIP}~\cite{kaffpa}, and \textsc{KaMinPar}~\cite{DeepMultilevelGraphPartitioning} achieve high-quality partitions but require memory proportional to the entire graph, making them impractical for billion-edge or rapidly evolving graphs. For modern applications that require the processing of complex graphs in real-time, streaming graph partitioners offer a resource-efficient alternative by processing nodes sequentially without storing the full graph in memory. However, this efficiency comes at the cost of reduced solution quality, as streaming methods must make placement decisions with limited structural information.
	
	The need for high-quality resource-efficient streaming partitioners is highly evident in machine learning pipelines, notably in distributed training of graph neural networks (GNNs), widely used in recommendation systems, natural language processing, biological discovery, and fraud detection~\cite{gnnsocialrec,gnncodesummarization,gnnfrauddetection,gnnprotein,gnnrecommendersystems}. GNN training is computationally intensive, involving repeated communication of high-dimensional feature vectors and storage of large intermediate embeddings across layers. As a result, GNNs are typically trained in distributed environments, where partitioning quality affects memory consumption, communication volume, and training efficiency~\cite{gnntrainingmayer}. The computational requirements of GNN workloads indicate a clear motivation for improving streaming partitioners across quality and efficiency.

	\begin{figure}[t]
		\centering
		\includegraphics[width=0.8\linewidth]{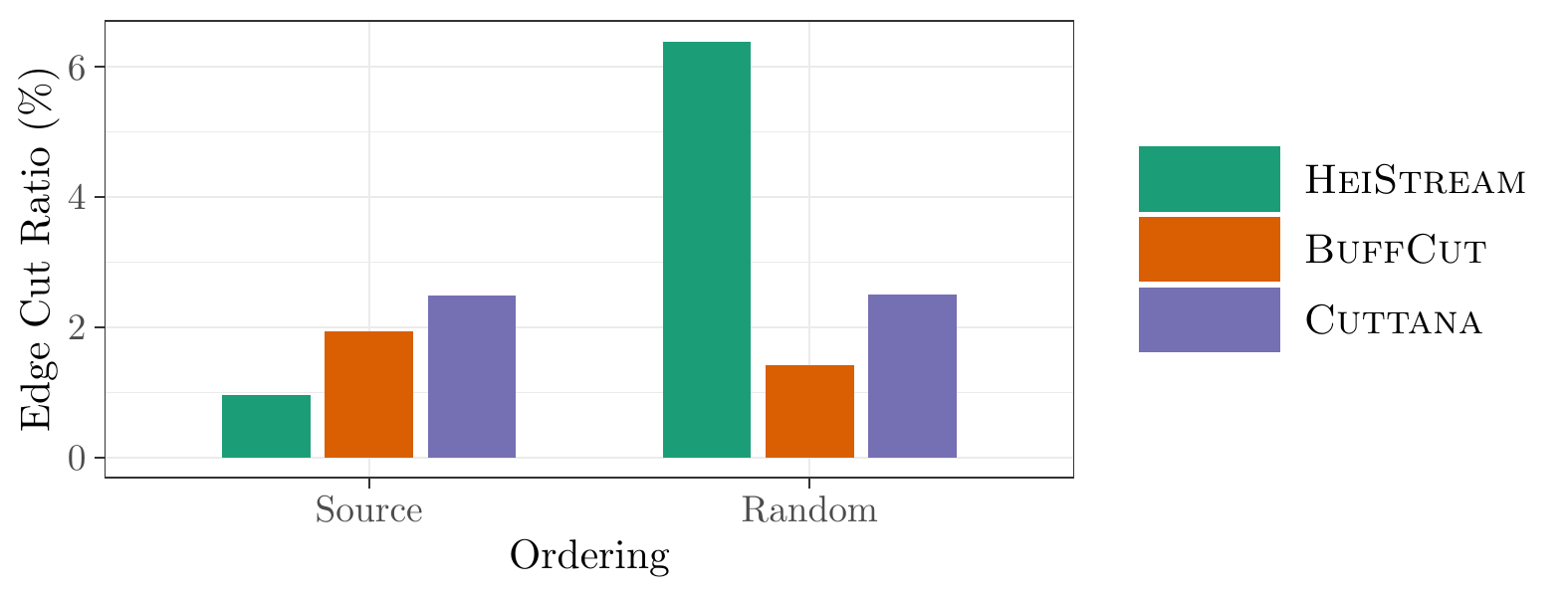}
		\caption{\textbf{Edge cut on source and random ordering.} The ratio of cut edges to total graph edges (\%) on \texttt{uk-2007-05} web graph ($k$=16) comparing source ordering (original node sequence in source file) to random ordering (independent random permutations of node IDs resulting in adversarial stream order) for \textsc{HeiStream}, \textsc{Cuttana} and our proposed \textsc{BuffCut}.}
		\label{fig:ukcut}
	\end{figure}
	
	Classical one-pass streaming heuristics such as \textsc{Fennel}~\cite{tsourakakis2014fennel}, while resource-efficient, lag in solution quality and are highly sensitive to stream order locality: when neighbors appear far apart in the stream, nodes are assigned with little structural context, resulting in substantial quality degradation. 
	Recent advances in buffered streaming attempt to bridge this gap. \textsc{HeiStream}~\cite{HeiStream} uses batch-wise multilevel partitioning to improve locality but remains vulnerable to adversarial input orders. For instance, on the \texttt{uk-2007} web graph ($k$=16), \textsc{HeiStream}'s edge cut degrades from 31.5M in the graph’s original node sequence, i.e. its source ordering, to 211.0M when the stream is randomized (Figure~\ref{fig:ukcut}). Conversely, \textsc{Cuttana}~\cite{cuttana} uses a prioritized buffer to handle poor orderings but relies on sequential assignments, missing the global refinement gains of batch-wise multilevel partitioning. While \textsc{Cuttana} improves the cut to 82.4M on the randomized instance, our proposed \ouralg achieves a significantly superior result of 46.7M, effectively recovering the structural information lost to randomization and approaching the quality achieved on the high-locality source ordering.
	In this paper, we contribute the following:
	
	\begin{itemize}
		\item{We introduce \ouralg, a buffered streaming partitioner that delays premature block assignments using a bounded priority buffer and incrementally forms batches with high internal locality for multilevel partitioning. \ouralg combines high-quality batch-wise partitioning with prioritized buffering for stream-order robustness.}
		
		\item{We introduce the Hub-Aware Assigned Neighbors Ratio buffer score that refines degree-aware prioritization to systematically balance node degree and neighborhood information.} 
		\item{We present a parallel implementation of \ouralg that accelerates end-to-end partitioning and restreaming passes for additional quality improvements.}
		
		\item{We share extensive experimental evaluation demonstrating that \ouralg outperforms existing streaming partitioners, achieving, in geometric means, 20.8\% fewer edge cuts while running 2.9$\times$ faster and using 11.3$\times$ less memory than \textsc{Cuttana}, the strongest prioritized buffering baseline.}
		
	\end{itemize}

	\section{Preliminaries}
	\label{sec:prelim}
	
	\subsection{Basic Concepts}
	\subparagraph*{Graphs and Graph Partitioning.}
	We consider an undirected graph $G=(V,E)$ with $n = |V|$ nodes and $m = |E|$ edges. Each node $v \in V$ has a weight $c(v) \in \MdR_{\ge 0}$, and each edge $e \in E$ has a weight $\omega(e) \in \MdR_{>0}$. Weight functions are extended additively to sets. We denote the neighborhood of $v$ as $N(v) = \{ u \in V \mid (u,v) \in E \}$, the degree as $d(v) = |N(v)|$, and the maximum degree in $G$ as $\Delta$.
	Given an integer $k \ge 1$, a \emph{$k$-partition} of a graph $G=(V,E)$ divides the node set into $k$ disjoint blocks $V_1,\ldots,V_k$ such that $\bigcup_{i=1}^k V_i = V$. The objective is to minimize the \emph{edge cut} $\omega(E_{\mathrm{cut}})$, where $E_{\mathrm{cut}} \coloneqq \{ (u,v) \in E \mid u \in V_i, v \in V_j, i \neq j \}$, subject to the balance constraint $c(V_i) \le L_{\max} \coloneqq \lceil (1+\varepsilon)\frac{c(V)}{k} \rceil$ for a given imbalance parameter~$\varepsilon \ge 0$. Minimizing the cut reduces inter-block communication, while the balance constraint ensures uniform load distribution in parallel systems.
	
	\subparagraph*{Streaming Models.}
	In the \emph{node-streaming} model, nodes $v_1, \dots, v_n$ and their incident edges arrive sequentially for processing. We distinguish three variations.
	\begin{itemize}
		\item \textbf{One-pass}: Nodes are assigned to a block immediately upon arrival based on a scoring function that favors blocks with existing neighbors while penalizing imbalance.
		\item \textbf{Buffered}: Nodes are held in sliding buffers of fixed capacity which expose partial structural information that allows for more informed assignments.
		\item \textbf{Restreaming}: The input is processed in multiple passes, allowing assignments to be refined based on global state from previous passes.
	\end{itemize}

	\subparagraph*{Stream Order Locality.}
	The performance of streaming heuristics is highly sensitive to the stream order, defined as a permutation $S = (v_1, \dots, v_{n})$ of $V$. The \emph{source ordering} is the order in which the graph is stored in the source file. This often exhibits high locality such as when nodes are generated through clustering or traversal-based schemes. We quantify stream locality using the \emph{Neighbor to Neighbor Average ID Distance (AID)}~\cite{esfahanilocality}. For a node $v$ with neighbors $N(v) = \{u_1, \dots, u_{d(v)}\}$ sorted by their position in the stream, $AID$ is defined as:
	\begin{equation}
		\label{aid}
		AID_v = \frac{1}{d(v)} \sum_{i = 2}^{d(v)} |u_i - u_{i-1}|
	\end{equation}
	Graph-level locality is the mean $AID_v$ across all $v \in V$. Lower values indicate higher locality, where neighbors appear in close proximity within the stream.
	
	\subsection{Related Work}
	\label{subsec:rel_work}
	
	We refer readers to recent surveys for general graph partitioning~\cite{SPPGPOverviewPaper,more_recent_advances_hgp} and focus here on streaming partitioners.
	Early streaming approaches are based on one-pass algorithms like \textsc{LDG}~\cite{stanton2012streaming} and \textsc{Fennel}~\cite{tsourakakis2014fennel} which greedily assign nodes upon arrival. \textsc{Fennel} assigns node~$v$ to the block $V_i$ maximizing $g(v, V_i) = |N(v) \cap V_i| - f(|V_i|)$, where $f(|V_i|) = \alpha \gamma |V_i|^{\gamma-1}$ is an additive penalty to enforce balance. While computationally efficient, these methods make poor assignment decisions as they lack global context. Restreaming approaches~\cite{nishimura2013restreaming} mitigate this by performing multiple passes to refine assignments using state from previous iterations.
	
	Buffered streaming partitioners retain a subset of the graph in memory to exploit partial structural information. \textsc{HeiStream}~\cite{HeiStream} iteratively loads batches of nodes together with their incident edges and assigns blocks using a weighted variant of the \textsc{Fennel} objective within a multilevel partitioning scheme. While batch-wise assignment significantly improves partition quality, \textsc{HeiStream} remains vulnerable to adversarial or low-locality stream orders~\cite{cuttana}.
	\textsc{Cuttana}~\cite{cuttana} addresses order sensitivity through a two-phase prioritized buffering scheme. In phase 1, nodes are temporarily stored in a priority queue and ranked by the \emph{Cuttana Buffer Score (CBS)} to \hbox{avoid premature assignments:}
	\begin{equation}
		\label{score:cbs}
		CBS(v) = \frac{d(v)}{D_{\max}} + \theta \cdot \frac{\sum_{i=1}^k |N(v) \cap V_i|}{d(v)}
	\end{equation}
	where $D_{\max}$ is a degree threshold. The first term favors high-degree nodes, while the second emphasizes the fraction of already assigned neighbors. When the buffer reaches capacity, the top node is evicted and assigned using a modified \textsc{Fennel} function. In Phase~2, \textsc{Cuttana} applies a refinement algorithm where nodes are grouped into sub-partitions, and coarse-grained trades between partitions are performed.  While robust to adversarial ordering, \textsc{Cuttana} assigns nodes sequentially upon buffer eviction, failing to exploit the locality-capturing potential of multilevel batch partitioning. 
	
	\section{BuffCut: Prioritized Buffered Streaming Partitioning}
	
	\begin{algorithm}[t]
		\caption{\ouralg (sequential, one-pass): core streaming loop with prioritized buffering and batch-wise multilevel assignment.}
		\label{alg:buffcut}
		
		\SetKwInOut{Input}{Input}
		\SetKwInOut{Output}{Output}
		\SetKwProg{Proc}{Procedure}{}{end}
		\SetKwFunction{PartitionBatch}{PartitionBatch}
		
		\Input{Graph $G=(V,E)$; blocks $k$; buffer size $\mathcal{Q}_{\max}$; batch size $\delta$; hub threshold $D_{\max}$; buffer score $s(\cdot)$}
		\Output{Assignment $\textnormal{block}(\cdot)$}
		
		init buffer $\mathcal{Q}\gets\emptyset$, batch $\mathcal{B}\gets\emptyset$, $\textnormal{block}(v)\gets\bot~\forall v$\;
		
		\ForEach{streamed node $v$ with neighbor list $N(v)$}{
			\uIf{$d(v)>D_{\max}$}{
				$\textnormal{block}(v)\gets \textsc{Fennel}(v,k)$\tcp*{assign hubs immediately}
				\ForEach{$u\in N(v)\cap \mathcal{Q}$}{
					$\mathcal{Q}$.\textsc{IncreaseKey}$(u,s(u))$\tcp*{update scores}
				}
			}\Else{
				$\mathcal{Q}$.\textsc{Insert}$(v,s(v))$\tcp*{buffer node}
			}
			
			\While{$|\mathcal{Q}|=\mathcal{Q}_{\max}$ \textbf{and} $|\mathcal{B}|<\delta$}{
				$u\gets\mathcal{Q}.\textsc{ExtractMax}()$\tcp*{evict top node}
				add $u$ to $\mathcal{B}$\;
				\ForEach{$w\in N(u)\cap \mathcal{Q}$}{
					$\mathcal{Q}$.\textsc{IncreaseKey}$(w,s(w))$\;
				}
			}
			
			\If{$|\mathcal{B}|=\delta$}{
				\PartitionBatch{$\mathcal{B},\textnormal{block}(\cdot),k$}\tcp*{partition and clear batch}
			}
		}
		
		\While{$\mathcal{Q}\neq\emptyset$}{
			add $\mathcal{Q}.\textsc{ExtractMax}()$ to $\mathcal{B}$\tcp*{flush}
			\If{$|\mathcal{B}|=\delta$}{
				\PartitionBatch{$G,\mathcal{B},\textnormal{block}(\cdot),k$}\;
			}
		}
		\If{$|\mathcal{B}|>0$}{
			\PartitionBatch{$G,\mathcal{B},\textnormal{block}(\cdot),k$}\tcp*{final}
		}
		
		\Proc{\PartitionBatch{$\mathcal{B},\textnormal{block}(\cdot),k$}}{
			$G_{\mathcal{B}}\gets \textsc{BuildBatchModel}(\mathcal{B},\textnormal{block}(\cdot),k)$\tcp*{model graph}
			$\textnormal{block}(\mathcal{B})\gets \textsc{MLPartition}(G_{\mathcal{B}},k)$\tcp*{multilevel partition}
			$\mathcal{B}\gets\emptyset$\tcp*{clear}
		}
		
	\end{algorithm}
	
	Streaming partitioners are attractive for large-scale and frequently updating graphs because they operate within strict memory constraints and process the input online. However, they often yield lower quality under unfavorable stream orderings as early assignments are made with limited structural context. We present \ouralg, a buffered streaming partitioner that achieves high quality and reduces order sensitivity, while remaining more resource-efficient than in-memory approaches, by combining (i) prioritized buffering for high-locality batch formation and (ii) batch-wise multilevel assignment on a compact model graph.
	
	\subsection{Overall Algorithm}

\begin{figure}[t]
		\centering
		\includegraphics[width=1.0\linewidth]{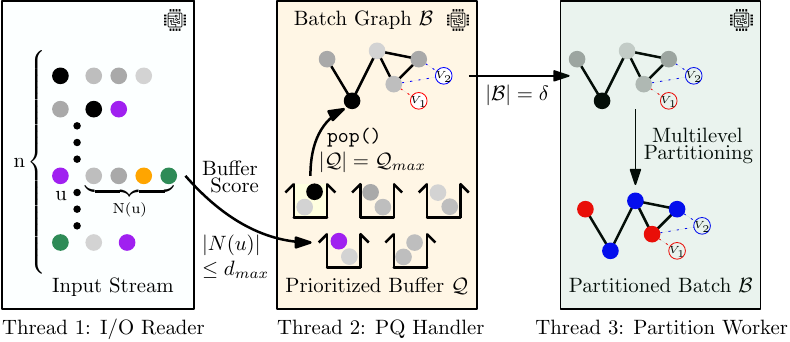}
		\caption{\textbf{Overview of \ouralg and its parallel pipeline.} Nodes arrive as a stream (left). Low-degree nodes are scored and inserted into a bounded prioritized buffer $\mathcal{Q}$, while hubs bypass the buffer. When $|\mathcal{Q}|=\mathcal{Q}_{\max}$, the highest-priority node is evicted to incrementally grow a batch $\mathcal{B}$ (middle); once $|\mathcal{B}|=\delta$, \ouralg constructs the batch model graph and partitions it via multilevel refinement, committing assignments (right). After committing, the batch $\mathcal{B}$ is cleared and construction resumes with the next evictions. The three stages are overlapped using an I/O reader, buffer handler, and partition worker.}
		\label{fig:overall}
	\end{figure}

	\ouralg processes the input graph $G$ as a node stream under explicit, user-configurable memory bounds. It uses \emph{buffering} to remain robust to stream order: rather than assigning each streamed node immediately, \ouralg delays placement until sufficient neighborhood evidence becomes available. Subsequently, several nodes are partitioned simultaneously using a multilevel partitioning scheme.
	The algorithm maintains two working sets with distinct purposes, terminology and separately configurable sizes: a bounded priority \emph{buffer} $\mathcal{Q}$ of capacity~$\mathcal{Q}_{\max}$ that stores deferred nodes and continuously reorders them by importance, and a \emph{batch} $\mathcal{B}$ of target size $\delta$ whose nodes are assigned jointly in a multilevel partitioning scheme. 
	Upon arrival of a node $v$, \ouralg treats high- and low-degree nodes differently. High-degree nodes ($d(v)>D_{\max}$) are assigned immediately using \textsc{Fennel}, which establishes stable anchors for subsequent decisions. All remaining nodes are inserted into $\mathcal{Q}$, implemented as a bucket-based max-priority queue keyed by a buffer score~$s(v)$ that quantifies how informative~$v$'s current neighborhood is with respect to already assigned nodes (Section~\ref{sec:scores}).
	Whenever the buffer reaches capacity ($|\mathcal{Q}|=\mathcal{Q}_{\max}$), \ouralg evicts the highest-priority buffered node into the active batch $\mathcal{B}$ until $|\mathcal{B}|=\delta$. Batch construction is incremental to allow frequent re-prioritization: as nodes are admitted to $\mathcal{B}$ (or assigned immediately as hubs), the buffer scores of their buffered neighbors are updated, so newly revealed neighborhood information immediately influences subsequent evictions and promotes high internal locality within $\mathcal{B}$ (Section~\ref{sec:prioritized_batched_streaming}). Once the batch is full, \ouralg constructs a compact batch model graph $G_{\mathcal{B}}$ and assigns all nodes in $\mathcal{B}$ via multilevel partitioning (Section~\ref{sec:multilevel}). The batch is cleared and the process continues until all nodes are assigned. Algorithm~\ref{alg:buffcut} provides pseudocode for a single sequential pass and Figure~\ref{fig:overall} gives an overview. \ouralg supports restreaming for additional refinement and a parallel \hbox{implementation for speed (Section~\ref{sec:par_re}).}

	\subsection{Prioritized Buffered Streaming}
	\label{sec:prioritized_batched_streaming}
	
	We describe buffer and batch construction in \ouralg, highlighting two key improvements over \textsc{Cuttana}~\cite{cuttana}, which uses degree-aware buffering to reorder nodes for sequential assignment with subsequent refinement (Section~\ref{subsec:rel_work}). Firstly, \ouralg does not assign nodes immediately upon buffer eviction; instead,
	evicted nodes are accumulated into batches that are assigned jointly using multilevel partitioning, yielding higher-quality solutions.
	Moreover, \ouralg uses a more efficient priority queue implementation, which reduces overall buffer maintenance time.
	
	As observed by Hajidehi et al.~\cite{cuttana}, low-degree nodes are highly prone to under-informed placement during streaming. \ouralg therefore assigns high-degree nodes with $d(v) > D_{\max}$ immediately using \textsc{Fennel}, while buffering nodes with $d(v)\le D_{\max}$ in a priority queue $\mathcal{Q}$. Immediate assignment of high-degree hubs establishes stable anchors for neighboring nodes while preventing the buffer from being dominated by memory-intensive nodes that would trigger excessive score updates.
	
	When the buffer $\mathcal{Q}$ reaches its maximum size $\mathcal{Q}_{\max}$, \ouralg extracts the top node from~$\mathcal{Q}$ and appends it to the active batch $\mathcal{B}$ until $|\mathcal{B}|=\delta$ (Algorithm~\ref{alg:buffcut}). Instead of evicting~$\delta$ nodes from $\mathcal{Q}$ at once, batch construction is incremental to enable frequent updates to node order within the buffer. As soon as a node $u$ is admitted to a batch, it is treated as assigned for the purpose of buffer score computation, though its block assignment is deferred until the batch model graph is partitioned. Consequently, the buffer scores of its buffered neighbors~$v \in N(u) \cap \mathcal{Q}$ are updated immediately, allowing newly available neighborhood information to influence subsequent ordering decisions, with the same update applied when a hub is assigned immediately. This tight feedback loop promotes the grouping of structurally related nodes into the same batch, even when they appear far apart in the input stream.
	
	\begin{algorithm}[t]
		\caption{Bucket Priority Queue operations for efficient buffer score updates.}
		\label{alg:bucket_pq_universal}
		\SetKwProg{Proc}{Procedure}{}{end}
		
		\BlankLine
		\textbf{State:}
		Array of dynamic arrays $\mathsf{Buckets}[0 \dots B-1]$\;
		Location map $\mathsf{L}[v] = (b, p)$ (bucket $b$, position $p$)\;
		Top pointer $\rho = \max\{b : \mathsf{Buckets}[b] \neq \emptyset\}$\;
		\textbf{Key:} discretize score $s(v)$ as $\mathrm{idx}(v)=\min\{\mathrm{round}(s(v)\cdot discFactor),\,B-1\}$\;
		
		\Proc{\texttt{Insert}$(v,s(v))$\tcp*[f]{$O(1)$}}{
			$b\gets \mathrm{idx}(v)$\;
			$\mathsf{Buckets}[b].\textsc{Push}(v)$\;
			$\mathsf{L}[v]\gets (b,\ |\mathsf{Buckets}[b]|-1)$\;
			$\rho\gets \max(\rho,b)$\;
		}
		
		\Proc{\texttt{IncreaseKey}$(v,s(v))$}{
			$(b,p)\gets \mathsf{L}[v]$\;
			$x\gets \mathsf{Buckets}[b].\textsc{Pop}()$\tcp*{pop $O(1)$}
			\If{$p < |\mathsf{Buckets}[b]|$}{
				$\mathsf{Buckets}[b][p]\gets x$\tcp*{swap $O(1)$}
				$\mathsf{L}[x]\gets (b,p)$\;
			}
			\textsc{Insert}$(v,s(v))$\tcp*{$O(1)$}
		}
		
		\Proc{\texttt{ExtractMax}() $\to v$}{
			$v\gets \mathsf{Buckets}[\rho].\textsc{Pop}()$\tcp*{$O(1)$}
			\While{$\rho>0$ \textbf{and} $\mathsf{Buckets}[\rho]$ empty}{
				$\rho\gets \rho-1$\tcp*{rare worst-case $O(B)$}
			}
			\Return $v$\;
		}
		
	\end{algorithm}
	
	Frequent buffer score updates are required to re-prioritize neighbors during batch construction.
	Over a streaming pass, scanning the neighborhoods of all nodes results in a total of $\mathcal{O}(m)$ updates. 
	Specifically, each edge contributes to at most one such update event when one endpoint is admitted to a batch, while the other endpoint remains buffered.
	In \textsc{Cuttana}, the buffer is implemented using a \texttt{set}-based priority queue, imposing a logarithmic penalty of $\mathcal{O}(\log \mathcal{Q}_{\max})$ per update due to tree rebalancing. 
	To eliminate this bottleneck, \ouralg implements $\mathcal{Q}$ as a bucket priority queue that exploits the bounded range and monotonicity of buffer scores to achieve amortized constant time updates. 
	We first map continuous buffer scores to a finite set of priority levels.
	Let $s(v) \in [0, S_{\max}]$ denote the buffer score of node $v$. We discretize these scores into $B$ integer buckets using a scaling factor $discFactor$, mapping each node to a bucket index \hbox{$\mathrm{idx}(v)=\min\{\mathrm{round}(s(v)\cdot discFactor),\,B-1\}$}.
	We store $\mathcal{Q}$ as an array of $B$ dynamic arrays $\mathsf{Buckets}[0 \dots B-1]$ together with a location map $L[v]=(b,p)$ indicating the current bucket $b$ and position $p$ of each buffered node, and a top pointer $\rho$ to the highest non-empty bucket (Algorithm~\ref{alg:bucket_pq_universal}).
	
	During batch construction, buffer scores are monotonic non-decreasing, so all priority updates are \texttt{IncreaseKey} operations.
	We implement \texttt{IncreaseKey} in $\mathcal{O}(1)$ amortized time via a pop-and-swap operation followed by an \texttt{Insert}: to move a node $v$ to a higher-index bucket, we swap $v$ with the last element in its current bucket, update the location map $\mathsf{L}$ for the swapped element, and append $v$ to the target bucket.
	\texttt{Insert} is an append plus constant-time updates to $\mathsf{L}$ and $\rho$, and thus also runs in $\mathcal{O}(1)$ amortized time with dynamic arrays.
	\texttt{ExtractMax} removes an element from $\mathsf{Buckets}[\rho]$ and then decreases $\rho$ until the next non-empty bucket is found.
	This scan can traverse at most $B$ buckets, and hence \texttt{ExtractMax} takes $\mathcal{O}(B)$ worst-case time. 
	In our setting, however, this worst case is rare. Since buffer scores increase gradually, and $B$ is fixed to a constant that is orders of magnitude smaller than $m$, buckets are dense and the top pointer typically only moves locally.  
	
	\subsection{Buffer Scoring Functions}
	\label{sec:scores}
	\begin{figure}[t]
		\centering
		\resizebox{0.7\linewidth}{!}{%
			\begin{minipage}{\linewidth}
				\centering
				\begin{subfigure}[t]{0.49\linewidth}
					\centering
					\includegraphics[width=\linewidth]{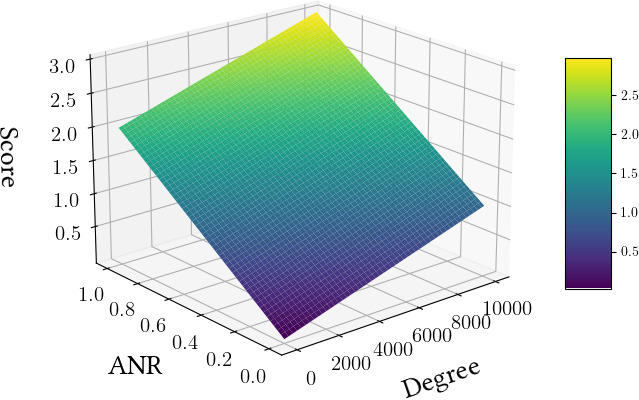}
					\caption{CBS ($\theta = 2$)}
				\end{subfigure}\hfill
				\begin{subfigure}[t]{0.49\linewidth}
					\centering
					\includegraphics[width=\linewidth]{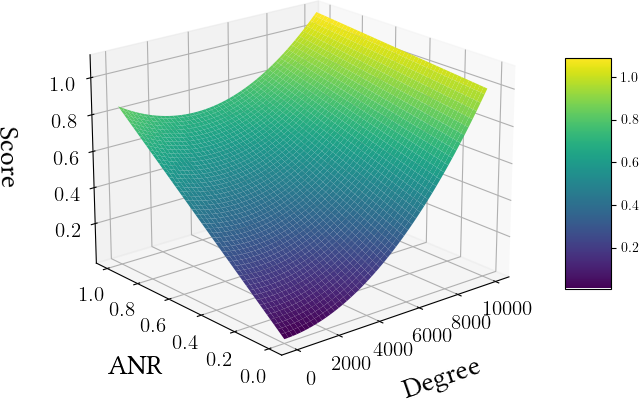}
					\caption{HAA ($\theta = 0.75, \beta = 2$)}
				\end{subfigure}
			\end{minipage}
		}
		\caption{\textbf{Heatmap visualizations of CBS and HAA} as a function of degree (x-axis)
			and assigned-neighbor ratio (y-axis). Bright colors indicate high scores, which lead to
			earlier eviction from the buffer.}
		\label{fig:cbs_haa_heatmap}
	\end{figure}
	
	Our algorithm is parameterized by a buffer score $s(\cdot)$ that determines the eviction order from~$\mathcal{Q}$. We define four buffer scores that quantify how informative a node’s neighborhood is and thus determine its priority in the buffer, shaping batch composition, locality, and ultimately partition quality. Section~\ref{sec:exp_buffer_scores} shows our experimental evaluation of these scores (including \textsc{Cuttana}’s). As validated in our experiments, our proposed Hub-Aware Assigned Neighbors Ratio improves on \textsc{Cuttana}'s buffer score. Let $v_t$ be the node streamed at time~$t$, with neighborhood $N(v_t)$ and degree $d(v_t)=|N(v_t)|$. Let $V_1,\dots,V_k$ be the current blocks and $\text{block}(u)$ the assigned block of any already placed node $u$. For degree normalization, we define $\hat d(v_t)=d(v_t)/D_{\max}$, where $D_{\max}$ is the threshold for buffering (Section~\ref{sec:prioritized_batched_streaming}). For all scores, larger values indicate higher buffer priority.
	
	\begin{enumerate}
		\item \textbf{Assigned Neighbors Ratio (ANR).}
		ANR measures the fraction of a node’s neighbors that have
		already been assigned to some partition:
		\begin{equation}
			\label{score:anr}
			\mathrm{ANR}(v_t) =
			\frac{\sum_{i=1}^k |N(v_t) \cap V_i|}{d(v_t)}
		\end{equation}
		
		\item \textbf{Hub-Aware Assigned Neighbors Ratio (HAA).}
		Building on the \emph{Cuttana Buffer Score} (Equation~\ref{score:cbs}), we introduce a refined parametric score in which the influence of ANR decreases with node degree, prioritizing hubs by degree and low-degree \hbox{nodes by ANR:}
		\begin{equation}
			\label{score:haa}
			\mathrm{HAA}(v_t) =
			\hat d(v_t)^{\beta}
			+ \theta \cdot \bigl(1 - \hat d(v_t)\bigr) \cdot \mathrm{ANR}(v_t)
		\end{equation}
		where $\beta \geq 1$ controls the shape of the degree contribution and $\theta$ balances
		the relative influence of neighborhood information.
		
		\item \textbf{Neighborhood Seen Score (NSS).}
		NSS factors in neighbors currently present in the buffer $\mathcal{Q}$. For a weighting parameter $\eta \in [0,1]$:
		\begin{equation}
			\label{score:nss}
			\mathrm{NSS}_{\eta}(v_t) =
			\frac{\sum_{i=1}^k |N(v_t) \cap V_i| + \eta \, |N(v_t) \cap \mathcal{Q}|}{d(v_t)}
		\end{equation}
		
		\item \textbf{Community-Majority Score (CMS).}
		CMS prioritizes nodes by the largest number of assigned neighbors in the same partition to reflect underlying community structure:
		\begin{equation}
			\label{score:cms}
			\mathrm{CMS}(v_t) =
			\max_{p \in \{1,\dots,k\}}
			\frac{|\{u \in N(v_t) : \text{block}(u) = p\}|}{d(v_t)}
		\end{equation}
	\end{enumerate}
	
	ANR (Eq.~\ref{score:anr}) measures available neighborhood assignment information: nodes with many already-assigned neighbors have more placement evidence and are evicted earlier. However, since ANR is normalized by $d(v_t)$, it systematically favors low-degree nodes and delays hubs, even though hubs serve as structural anchors for subsequent placements.
	The \emph{Cuttana Buffer Score} (CBS, Eq.~\ref{score:cbs}) addresses this by adding an explicit degree term, but combines it linearly with ANR, thus prioritizing a node only once both degree and ANR are large.
	HAA (Eq.~\ref{score:haa}) decouples these effects by decreasing the weight of ANR with normalized degree: the degree term $\hat d(v_t)^\beta$ prioritizes high-degree nodes as anchors irrespective of available neighborhood information, while the ANR term is emphasized for low-degree nodes via $(1-\hat d(v_t))\cdot \mathrm{ANR}(v_t)$. Thus, HAA promotes two desirable classes of early evictions: (i) low-degree nodes once their neighborhood evidence becomes reliable, and (ii) hubs even when their assigned-neighbor ratio is still small. Figure~\ref{fig:cbs_haa_heatmap} visualizes this separation: HAA assigns high priority to both ``low-degree/high-ANR'' and ``high-degree/low-ANR'' regions, whereas CBS concentrates priority near the diagonal where both signals are high.

	\subsection{Batch-Wise Multilevel Partitioning}
	\label{sec:multilevel}
	Batched nodes in $\mathcal{B}$ and their incident edges define a \emph{batch model graph} that is partitioned with the multilevel scheme used in \textsc{HeiStream}~\cite{HeiStream}, summarized here. The model subgraph is augmented with $k$ auxiliary nodes $a_i$, each representing a block $i$ of the partition. 
	For a node $v$ and block $i$, we add an auxiliary edge $(v,a_i)$ if $v$ has at least one neighbor assigned to block $i$; its weight equals the number of such neighbors. Multilevel partitioning on the model graph proceeds: nodes are iteratively contracted to produce a hierarchy of smaller graphs, an initial partition is computed on the coarsest level using a weighted variant of the \textsc{Fennel} objective, and the solution is successively refined during uncoarsening~\cite{HeiStreamEdge,HeiStream}. After multilevel partitioning, block assignments on the model subgraph are mapped back to the global partition. In \textsc{HeiStream}, this mapping is trivial because nodes are streamed in input order, so local IDs are offsets from the batch’s first global ID. In \ouralg, nodes are buffered and processed out of order, requiring explicit local–to-global mappings. 
	
	\subsection{Parallelization and Restreaming}
	\label{sec:par_re}
	
	\subparagraph*{Parallelization.}
	We implement a parallel version of \ouralg that overlaps I/O, buffer management, and partitioning in three concurrent threads using bounded lock-free queues.
	Thread~1 (\emph{I/O Reader}) parses the stream and pushes \texttt{ParsedLine} objects into \texttt{input\_queue}.
	Thread~2 (\emph{Priority Queue Handler}) pops from \texttt{input\_queue}, computes buffer scores, maintains the priority buffer, and emits single-node or batch \texttt{PartitionTask}s into \texttt{partition\_task\_queue}.
	Thread~3 (\emph{Partitioning Worker}) executes tasks (immediate assignment or batch multilevel partitioning) and commits the resulting assignments.
	Because partitioning tasks overlap with continued buffering, the parallel schedule can yield slightly different batch composition than the sequential run; to keep scoring consistent, nodes are treated as \emph{assigned} for buffer scoring as soon as their task is enqueued.
	\subparagraph*{Restreaming.} To improve solution quality, \ouralg supports restreaming with a configurable number of passes. The first pass operates as described above; subsequent passes refine the existing partition without buffering or prioritization. Since all nodes are assigned after the first pass, neighborhood-based ordering is no longer meaningful; nodes are processed sequentially and repartitioned using batch-wise multilevel refinement. 
	
	\section{Experimental Analysis}
	\label{sec:experimental}
	We evaluate \ouralg~on a diverse set of real-world and synthetic graphs and \hbox{answer the following:}
	\begin{itemize}
		\item \textbf{RQ1:} How effective are prioritized buffering and batch-wise partitioning in improving partition quality?
		
		\item \textbf{RQ2:} What is the impact of restreaming and parallelization on solution quality, runtime and memory consumption?
		
		\item \textbf{RQ3:} How does \ouralg compare to state-of-the-art buffered streaming partitioners in partition quality, memory and runtime?
	\end{itemize}
	
	We first describe our baselines, setup, instances, and evaluation methodology, then address our research questions through parameter studies on a Tuning Set and comparative analysis against state-of-the-art partitioners on a larger Test Set.
	
	\subparagraph*{Setup and Reproducibility.} 
	All experiments were run on a dedicated machine with an \texttt{AMD EPYC 9754} CPU (128 cores, 256 threads, base frequency 2.25GHz), 755 GiB of DDR5 main memory, and an L2/L3 cache of 128 MiB / 256 MiB. The system also features an 894 GB NVMe solid-state drive and runs Ubuntu 22.04.4 LTS with Linux kernel version 5.15.0-140. 
	We implemented \ouralg in \texttt{C++} within \textsc{HeiStream}\footnote{The implementation is available as open source at \url{https://github.com/KaHIP/HeiStream}.} and compiled it with full optimization enabled (\texttt{-O3}). \ouralg uses $\mathit{discFactor}$ = $1\,000$, and $D_{\max} = 10\,000$ for all experiments.
	For comparison against the state-of-the-art, we obtained official implementations of \textsc{HeiStream}~\cite{HeiStream} and \textsc{Cuttana}~\cite{cuttana}. Here, \textsc{HeiStream} refers to the legacy implementation accompanying in its original paper, prior to the integration of \ouralg in it. 
	\textsc{HeiStream} requires a batch-size parameter~$\delta$ (default~32\,684); unless mentioned otherwise, we use $\delta = 1\,048\,576$ to ensure a fair comparison in terms of memory consumption.
	For \textsc{Cuttana}, we use the parallel implementation provided by the authors.
	To ensure that reported memory usage reflects algorithmic behavior rather than differences in file handling, we replace \textsc{Cuttana}'s default memory-mapped I/O with standard streaming access (\texttt{ifstream}), consistent with \textsc{HeiStream} and our implementation.
	\textsc{Cuttana} is evaluated using the parameters recommended in the publication: $D_{\max} = 1\,000$, a queue size of $10^6$ and a subpartition ratio $k'/k = 4\,096$ (referred to as \textsc{Cuttana4k}).
	We additionally evaluate a reduced configuration with $k'/k = 16$ (referred to as \textsc{Cuttana16}), which substantially lowers memory usage and runtime. 
	All experiments are run using GNU Parallel, with up to five concurrent instances by default and twelve for the KONECT ordering experiments.
	
	\begin{table*}[t]
		\centering
		\caption{\textbf{Graph Instances.}
			We use the Tuning Set for parameter studies and the Test Set for state-of-the-art comparisons, using three random stream orderings for each instance. Graphs with \textsuperscript{\dag} are also tested under KONECT’s first-appearance ordering~\cite{konect} to match the setup in~\cite{cuttana}.}
		\label{tab:datasets}
		\resizebox{1.0\textwidth}{!}{%
			\begin{tabular}{lrrl|lrrl}
				\toprule
				\multicolumn{4}{c|}{\textsc{Exploration \& Tuning Set}} &
				\multicolumn{4}{c}{\textsc{Test Set}} \\
				\midrule
				\textsc{Graph} & \textsc{$n$} & \textsc{$m$} & \textsc{Type} &
				\textsc{Graph} & \textsc{$n$} & \textsc{$m$} & \textsc{Type} \\
				\midrule
				
				\rowcolor{shade1}
				\texttt{coPapersDBLP} & 540\,486 & 15\,245\,729 & Cit. &
				\textsuperscript{\dag}\texttt{orkut} & 3\,072\,411 & 117\,185\,082 & Soc. \\
				
				\rowcolor{shade2}
				\texttt{soc-lastfm} & 1\,191\,805 & 4\,519\,330 & Soc. &
				\texttt{arabic-2005} & 22\,744\,080 & 553\,903\,073 & Web \\
				
				\rowcolor{shade1}
				\texttt{in-2004} & 1\,382\,908 & 13\,591\,473 & Web &
				\texttt{nlpkkt240} & 27\,933\,600 & 373\,239\,376 & Mat. \\
				
				\rowcolor{shade2}
				\texttt{Flan\_1565} & 1\,564\,794 & 57\,920\,625 & Mesh &
				\texttt{it-2004} & 41\,291\,594 & 1\,027\,474\,947 & Web \\
				
				\rowcolor{shade1}
				\texttt{G3\_Circuit} & 1\,585\,478 & 3\,037\,674 & Circ. &
				\textsuperscript{\dag}\texttt{twitter-2010} & 41\,652\,230 & 1\,202\,513\,046 & Soc. \\
				
				\rowcolor{shade2}
				\texttt{soc-flixster} & 2\,523\,386 & 7\,918\,801 & Soc. &
				\texttt{sk-2005} & 50\,636\,154 & 1\,810\,063\,330 & Web \\
				
				\rowcolor{shade1}
				\texttt{Bump\_2911} & 2\,852\,430 & 62\,409\,240 & Mesh &
				\texttt{com-Friendster} & 65\,608\,366 & 1\,806\,067\,135 & Soc. \\
				
				\rowcolor{shade2}
				\texttt{FullChip} & 2\,986\,999 & 11\,817\,567 & Circ. &
				\texttt{rgg26} & 67\,108\,864 & 574\,553\,645 & Gen. \\
				
				\rowcolor{shade1}
				\texttt{cit-Patents} & 3\,774\,768 & 16\,518\,947 & Cit. &
				\texttt{rhg1B} & 100\,000\,000 & 1\,000\,913\,106 & Gen. \\
				
				\rowcolor{shade2}
				\texttt{com-LJ} & 3\,997\,962 & 34\,681\,189 & Soc. &
				\texttt{rhg2B} & 100\,000\,000 & 1\,999\,544\,833 & Gen. \\
				
				\rowcolor{shade1}
				\texttt{Ljournal-2008} & 5\,363\,186 & 49\,514\,271 & Soc. &
				\textsuperscript{\dag}\texttt{uk-2007-05} & 105\,896\,555 & 3\,301\,876\,564 & Web \\
				
				\rowcolor{shade2}
				\texttt{italy-osm} & 6686493 & 7\,013\,978 & Road &
				\texttt{webbase-2001} & 118\,142\,155 & 854\,809\,761 & Web \\
				
				\rowcolor{shade1}
				\texttt{great-britain-osm} & 7\,733\,822 & 8\,156\,517 & Road &
				& & & \\
				
				\rowcolor{shade2}
				\texttt{uk-2002} & 18\,520\,486 & 261\,787\,258 & Web &
				& & & \\
				
				\bottomrule
			\end{tabular}
		}
	\end{table*}
	\subparagraph*{Datasets.}
	We evaluate on a diverse collection of real-world and synthetic benchmark graphs drawn from established repositories~\cite{benchmarksfornetworksanalysis,BMSB,BRSLLP,BoVWFI,KaGen,snap,nr-aaai15}.
	All instances have been used in prior work on graph partitioning and streaming algorithms~\cite{HeiStreamEdge,streamcpi,HeiStream}.
	We convert all graphs to the METIS format by removing self-loops and parallel edges, ignoring directions, and assigning unit weights to nodes and edges. Since \textsc{Cuttana} additionally requires node degrees to be explicitly provided, we convert compatible variants for it. 
	Our evaluation uses two instance groups: a \emph{Tuning Set} for parameter studies and a larger \emph{Test Set} for scalability and comparisons. 
	Table~\ref{tab:datasets} summarizes all instances. 
	
	Experiments on the Tuning and Test Sets use \emph{random stream orders}. 
	For each graph, we generate three independent random permutations of node IDs and report geometric means.
	To characterize stream locality, we measure their AID (Equation~\ref{aid}).
	Under source node orderings, the geometric mean AID is 37\,794 for the Tuning Set and 51\,104 for the Test Set.
	Under random orderings, these values increase to 219\,685 and 2\,581\,859, respectively, confirming that random orderings induce substantially lower locality and provide a challenging, order-agnostic evaluation setting.
	Graphs marked with \textsuperscript{\dag} in Table~\ref{tab:datasets} are additionally evaluated under the node order of these graphs as found in the KONECT repository~\cite{konect} (henceforth the \emph{KONECT ordering}) used by Hajidehi et al.~\cite{cuttana}. Concretely, when extracting these graphs from their source, KONECT renumbers nodes in \emph{first-appearance order} while scanning the edge list, which substantially reduces locality compared to the source ordering. We report results for this ordering to enable a direct comparison to the \textsc{Cuttana} publication.

	\subparagraph*{Methodology.}
	Unless stated otherwise, we enforce an imbalance of $\varepsilon = 3\%$.
	Tuning experiments are conducted with $k = 32$, and test experiments with $k \in \{4, 8, 16, 32, 64, 128, 256\}$. 
	We evaluate solution quality using the \emph{edge cut ratio} $\omega(E_{\mathrm{cut}})/\omega(E)$, alongside end-to-end runtime and peak memory (maximum resident set size). 
	To quantify how buffering impacts within-batch locality, we define the \emph{Internal Edge Ratio} (IER). For a batch $B \subseteq V$, let $E(B) \coloneqq \{(u,v) \in E : u,v \in B\}$ be the set of edges induced by $B$. We define the ratio as:
	\begin{equation}
		\label{eq:internal_edge_ratio}
		\mathrm{IER}(B) \coloneqq \frac{2\,\omega(E(B))}{\sum_{v \in B} d_{\omega}(v)}
	\end{equation}
	where $d_{\omega}(v)$ is the weighted degree of $v$. IER represents the fraction of incident edge weight contained entirely within $B$; we report the mean IER across all batches. For aggregate comparisons, we use geometric means and \emph{performance profiles}~\cite{pp}, which shows the fraction of instances where an algorithm's performance is within a factor $\tau$ of the best observed result.
	
	\begin{figure}[t]
		\centering
		\includegraphics[width=0.8\linewidth]{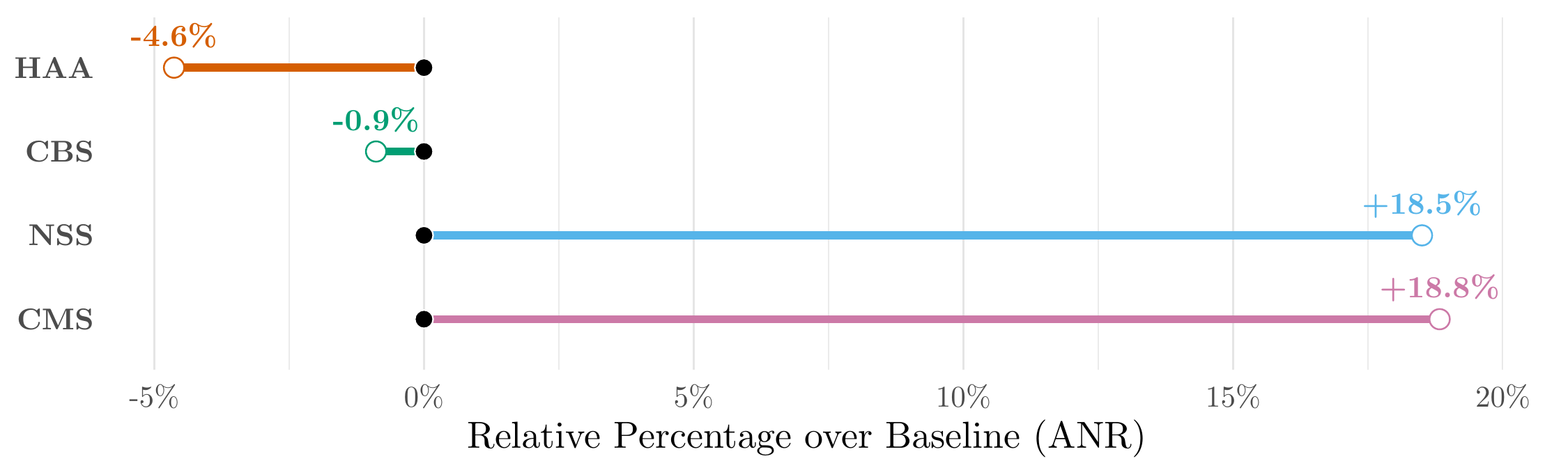}
		\caption{\textbf{Impact of buffer score on cut quality (Tuning Set, random order, $k=32$).} Geometric means of edge cut relative to ANR (lower is better).}
		\label{fig:score_comparison}
	\end{figure}

	\begin{figure}[t]
		\centering
		\begin{subfigure}[t]{0.49\linewidth}
			\centering
			\includegraphics[width=\linewidth]{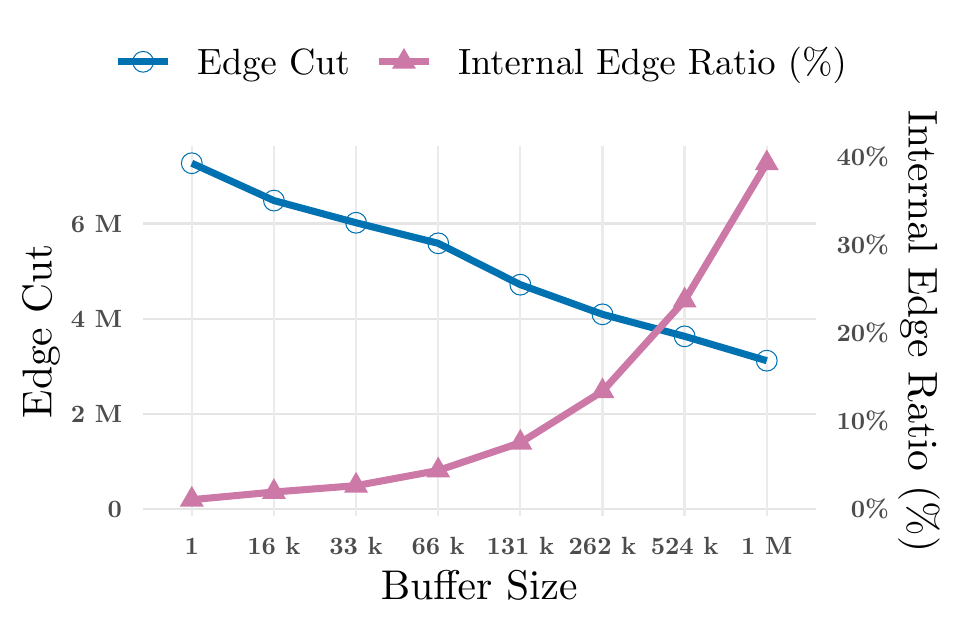}
			\caption{Edge Cut and Internal Edge Ratio}
			\label{fig:dual-edgecut-internal-buffer}
		\end{subfigure}\hfill
		\begin{subfigure}[t]{0.49\linewidth}
			\centering
			\includegraphics[width=\linewidth]{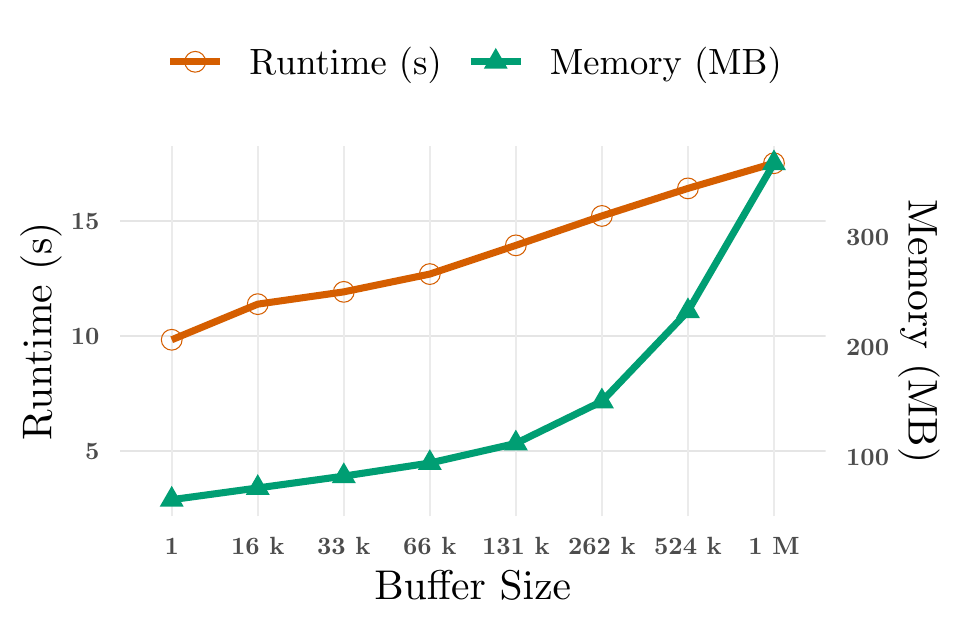}
			\caption{Runtime and Memory}
			\label{fig:dual-runtime-memory-buffer}
		\end{subfigure}
		
		\caption{\textbf{Effect of buffer size $\mathcal{Q}_{\max}$ (Tuning Set, random order, $k=32$).} Larger buffers increase within-batch locality (IER) and reduce cut at increased memory cost.}
		
		\label{fig:dual-sweep-buff}
	\end{figure}
	
	\begin{figure}[t]
		\centering
		\begin{subfigure}[t]{0.49\linewidth}
			\centering
			\includegraphics[width=\linewidth]{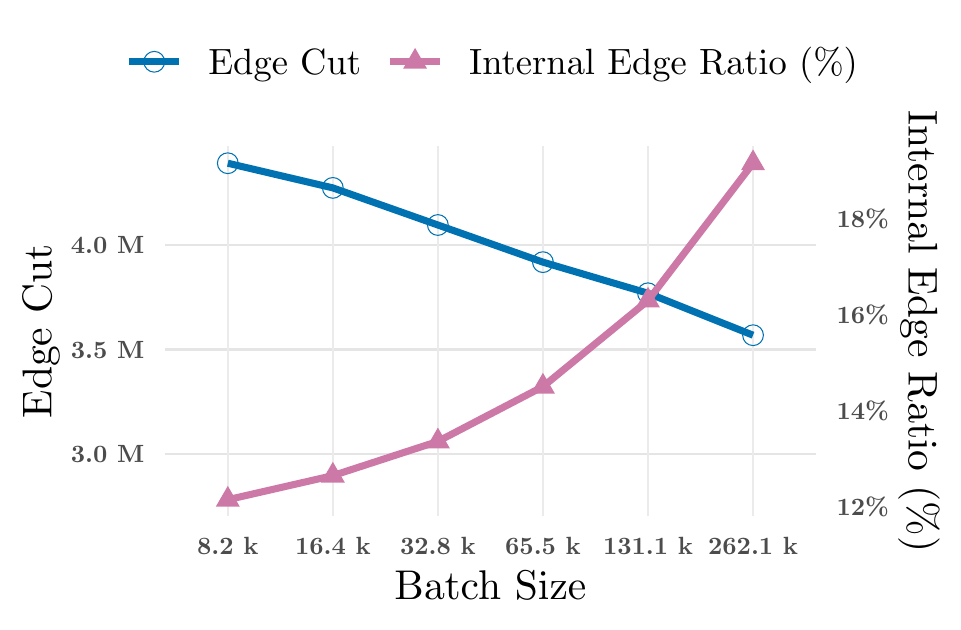}
			\caption{Edge Cut and Internal Edge Ratio}
			\label{fig:dual-edgecut-internal-batch}
		\end{subfigure}\hfill
		\begin{subfigure}[t]{0.49\linewidth}
			\centering
			\includegraphics[width=\linewidth]{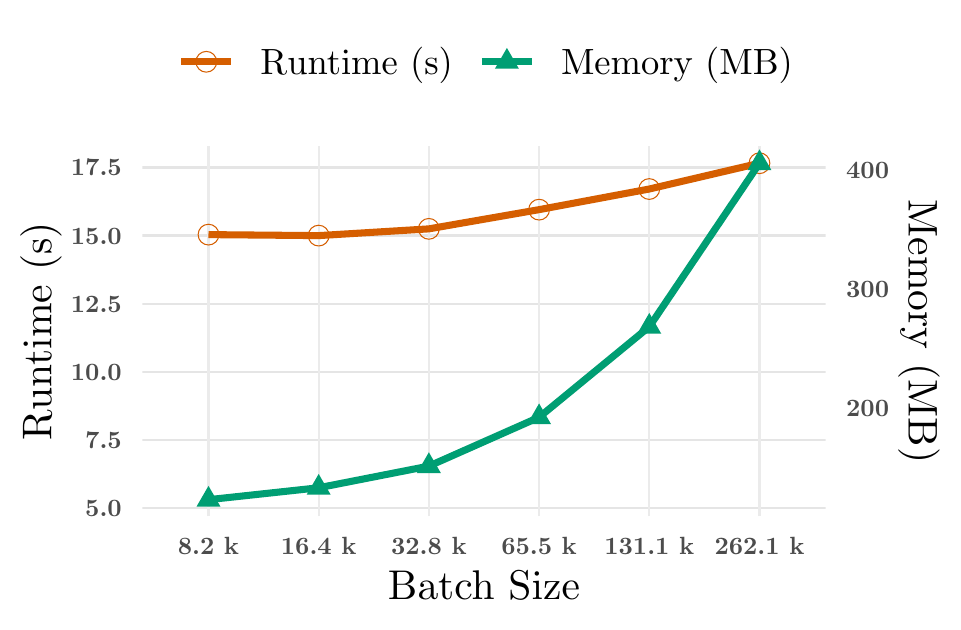}
			\caption{Runtime and Memory}
			\label{fig:dual-runtime-memory-batch}
		\end{subfigure}
		
		\caption{\textbf{Effect of batch size $\delta$ (Tuning Set, random order, $k=32$).} Larger batches improve cut but increase memory due to larger induced model graphs.}
		
		\label{fig:dual-sweep-batch}
	\end{figure}
	
	\begin{table}[t]
		\centering
		\caption{\textbf{Parallelization and restreaming trade-offs (Tuning Set, random order, $k = 32$).} Geometric means of edge cut ratio, runtime, and peak memory. Parallelization reduces runtime with negligible impact on cut; additional restreaming passes improve cut at higher runtime. \label{tab:seq_par_restream}}
		\resizebox{0.7\textwidth}{!}{%
			\begin{tabular}{llccc}
				\toprule
				& \textsc{Configuration} & \textsc{Edge cut (\%)} & \textsc{Runtime (s)} & \textsc{Memory (MB)} \\
				\midrule
				\rowcolor{shade1}
				& \textsc{Sequential} & 20.29 & 14.65 & 191.58 \\
				\rowcolor{shade2}
				\multirow{-2}{*}{\rotatebox{90}{\textsc{Par}}}
				& \textsc{Parallel}   & 20.48 & \textbf{7.85}  & 218.76 \\
				\midrule
				\rowcolor{shade1}
				& \textsc{1 stream}   & 20.29 & \textbf{15.60} & 191.81 \\
				\rowcolor{shade2}
				& \textsc{2 streams}  & 17.33 & 22.54 & 192.02 \\
				\rowcolor{shade1}
				& \textsc{3 streams}  & 16.71 & 29.50 & 192.17 \\
				\rowcolor{shade2}
				& \textsc{4 streams}  & 16.43 & 36.60 & 192.08 \\
				\rowcolor{shade1}
				\multirow{-5}{*}{\rotatebox{90}{\textsc{Restream}}}
				& \textsc{5 streams}  & \textbf{16.25} & 43.72 & 191.96 \\
				\bottomrule
			\end{tabular}
		}
	\end{table}
	
	\subsection{Prioritized Buffering and Batch-Wise Partitioning}
	\label{sec:exp_buffer_scores}
	We assess the buffer scoring functions from Section~\ref{sec:scores}. To address \textbf{RQ1}, we evaluate the effect of buffer size~$\mathcal{Q}_{\max}$, i.e., how many nodes are stored for prioritized ordering, and batch size~$\delta$, i.e., how many nodes are partitioned together in the multilevel scheme.
	
	\subparagraph*{Buffer Scoring Functions.}
	We evaluate buffer scoring functions with respect to partition quality, fixing $\mathcal{Q}_{\max} = 262\,144$ and $\delta = 32\,768$ (Figure~\ref{fig:score_comparison}). Our proposed \emph{Hub-Aware Assigned Neighbors Ratio} (HAA, with $\beta=2, \theta=0.75$) consistently yields the highest partition quality, reducing edge cuts by 4.6\% compared to the \emph{Assigned Neighbors Ratio} (ANR) baseline. 
	
	While the \emph{Cuttana Buffer Score} (CBS) also improves upon ANR by 0.9\%, it is consistently outperformed by HAA. 
	Although both CBS and HAA incorporate degree information, HAA performs better because it more effectively balances degree-based prioritization with neighborhood information, as discussed in Section~\ref{sec:scores}. NSS and CMS perform poorly, increasing edge cuts by $> 18\%$  relative to ANR, indicating that incorporating buffered neighbors provides little reliable information for prioritization, and emphasizing local majority effects leads to premature assignments. We utilize HAA as the default scoring \hbox{function for \ouralg.}

	\subparagraph*{Buffer Size ($\mathcal{Q}_{\max}$).}
	We study the effect of buffer size $\mathcal{Q}_{\max}$ on quality and computational overhead, fixing batch size at $\delta = 32\,768$ (Figure~\ref{fig:dual-sweep-buff}). Increasing $\mathcal{Q}_{\max}$ improves solution quality and within-batch locality, as measured by the Internal Edge Ratio (IER), by allowing the algorithm to defer assignments until more neighborhood information is available.
	
	Disabling the priority buffer ($\mathcal{Q}_{\max}=1$) results in the highest edge cut and negligible within-batch locality. Even modest buffer sizes provide significant gains: a buffer of $32\,768$ nodes reduces edge cut by 17.2\% compared to the no-buffer baseline with minimal overhead. As $\mathcal{Q}_{\max}$ increases to $262\,144$, the IER grows from 1\% to 13.4\%, yielding a 43.7\% reduction in edge cut. At the maximum tested buffer size of $2^{20} \approx 10^6$ nodes, \ouralg captures 39.2\% of edges within batches, reducing the edge cut by 57.1\% compared to using no buffer ($\mathcal{Q}_{\max}=1$).
	Runtime increases moderately (1.8$\times$ from $\mathcal{Q}_{\max}=1$ to $10^6$), with a sharper, non-linear increase in memory consumption (Figure~\ref{fig:dual-runtime-memory-buffer}). 
	These results demonstrate that \ouralg effectively trades memory for partition quality. Moderate buffer sizes ($\mathcal{Q}_{\max} \approx 262\,144$) represent a ``sweet spot'', capturing significant locality and reducing edge cut by over 40\% while maintaining low memory and runtime overhead. This highlights the effectiveness of prioritized buffering in mitigating the effects of low-locality stream orders.
	
	\subparagraph*{Batch Size ($\delta$).}
	We analyze the sensitivity of solution quality and computational overhead to the batch size $\delta$, with $\mathcal{Q}_{\max}$ fixed at $262\,144$ (Figure~\ref{fig:dual-sweep-batch}). Increasing $\delta$ consistently improves partition quality, as larger subgraphs provide the multilevel scheme with richer structural context. Specifically, expanding the batch size from $8\,192$ to $262\,144$ yields a 18.7\% reduction in edge cut, driven by a steady increase in IER from 12\% to nearly 20\%, while runtime increases moderately ($1.2\times$) and the memory footprint grows near-linearly ($3.3\times$). 
	Overall, batch size presents a trade-off between improved partition quality and increased memory and runtime overhead, with moderate batch sizes offering a favorable balance.
	
	\begin{takeaway}
		\textbf{Takeaway.} Our proposed HAA score outperforms all evaluated buffer scores, including \textsc{Cuttana'}s. Prioritized buffering and batch-wise partitioning improve solution quality, up to 57.1\% and 20\% on average, respectively, for the largest buffer and batch size tested, with computational overhead controlled via the configurable \hbox{buffer and batch size. } 
		
	\end{takeaway}
	
	\subsection{Parallelization and Restreaming}
	
	To answer \textbf{RQ2}, we compare the parallel and restreaming implementations of \ouralg to its default sequential, single stream implementation (Table~\ref{tab:seq_par_restream}), setting $\mathcal{Q}_{max} = 262\,144$ and $\delta = 65\,536$. Parallel \ouralg yields a substantial 1.87$\times$ speedup over the sequential implementation with the same solution quality, at the cost of a modest 14.2\% increase in memory consumption. Our results show that restreaming with five passes improves partitioning quality up to 19.9\%, but is 2.8$\times$ slower than single-stream \ouralg. Using two streams provides the best trade-off between solution quality and runtime, reducing the edge cut by 14.6\%, while being only 1.44$\times$ slower than using one stream. Restreaming does not incur proportional slowdowns because subsequent passes omit buffering.
	
	\subsection{Comparison to State-of-the-Art}
	\begin{figure}[t]
		\centering
		\includegraphics[width=1\linewidth]{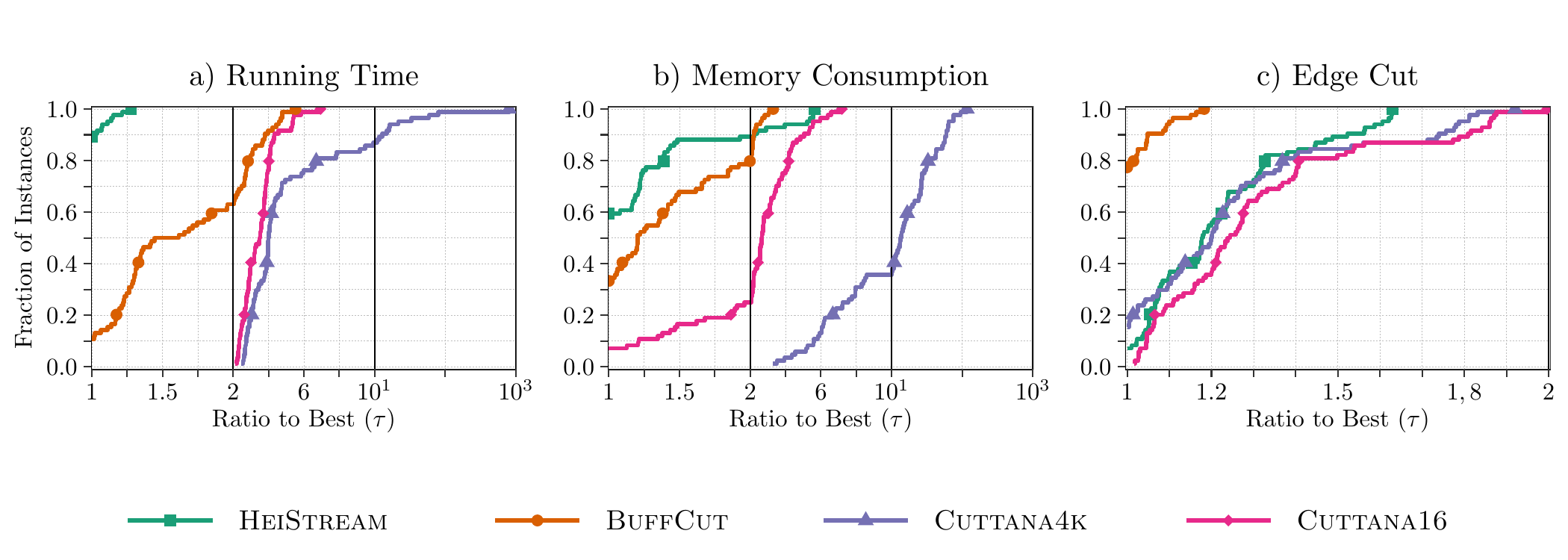}
		\caption{\textbf{Performance profiles on the Test Set (random order, $k\in\{4,\dots,256\}$).} Fraction of instances within factor $\tau$ of the best for edge cut, runtime, and peak memory.}
		\label{fig:test_set_pp}
	\end{figure}
	
	To answer \textbf{RQ3}, we compare parallelized \ouralg against all baselines on the Test Set under \emph{random} node orderings. In addition, for graphs marked with \textsuperscript{\dag} in Table~\ref{tab:datasets}, we report results under KONECT's reordering~\cite{konect} to match \textsc{Cuttana}'s experimental setup~\cite{cuttana}.
	
	\subparagraph*{Test Set Evaluation.} 
	We evaluate parallelized \ouralg ($\mathcal{Q}_{max} = 1\,048\,576$, $\delta = 65\,536$) against \textsc{HeiStream} and parallelized \textsc{Cuttana} on the Test Set under random stream orders (Figure~\ref{fig:test_set_pp}).
	\ouralg demonstrates clear dominance in solution quality, achieving the best edge cut on roughly 80\% of all instances. 
	While \textsc{HeiStream} is the most resource-efficient as it lacks prioritized reordering, achieving the best runtime and memory on 90\% and 60\% of instances, respectively, \ouralg provides a significantly more favorable quality-efficiency trade-off. Compared to \textsc{HeiStream}, \ouralg reduces the edge cut by 15.8\% on average with modest overheads of 1.8$\times$ in runtime and 1.09$\times$ in memory. 
	\ouralg consistently outperforms \textsc{Cuttana} across all metrics. 
	Both \textsc{Cuttana} variants are at least 2$\times$ slower on all instances; \textsc{Cuttana4k} is up to 1000$\times$ slower than the best baseline, while \ouralg is at worst 6$\times$ slower. Memory usage shows the same pattern: \textsc{Cuttana4k} always uses at least $3\times$ more memory than the best baseline and requires up to 100$\times$ more memory in the worst case. \textsc{Cuttana16} reduces this but still exceeds \ouralg on most instances. 
	Overall, \ouralg is the most memory-efficient in about 35\% of cases and always within approximately $3\times$ of the best baseline. Comparing geometric means across all instances and $k$ values, \ouralg achieves 20.8\% and 26.4\% fewer edge cuts than \textsc{Cuttana4k} and \textsc{Cuttana16}, respectively, while running 2.9$\times$ faster with 11.3$\times$ less memory than \textsc{Cuttana4k}, and 1.9$\times$ faster with~2$\times$ less memory than \textsc{Cuttana16}.

	\begin{table}[t]
		\centering
		\caption{\textbf{Quality--runtime comparison with $k=8$ and $5\%$ imbalance.}
			Edge cut ratio (\%) and running time (s); lower is better. Results are reported for original source ordering and their KONECT-ordering counterparts~\cite{konect}.
			Restreaming variants (one additional pass) are separated by a thick rule. Best values per column are highlighted.}
		\label{tab:k8_quality_runtime}
		
		\resizebox{1.0\textwidth}{!}{%
			\begin{tabular}{lcccccccc}
				\toprule
				& \multicolumn{2}{c}{\texttt{twitter}} 
				& \multicolumn{2}{c}{\texttt{uk-2007-05}} 
				& \multicolumn{2}{c}{\texttt{orkut}} 
				& \multicolumn{2}{c}{\texttt{uk-2002}} \\
				\cmidrule(lr){2-3} \cmidrule(lr){4-5} \cmidrule(lr){6-7} \cmidrule(lr){8-9}
				\textsc{Algorithm}
				& \textsc{Source} & \textsc{KONECT}
				& \textsc{Source} & \textsc{KONECT}
				& \textsc{Source} & \textsc{KONECT}
				& \textsc{Source} & \textsc{KONECT} \\
				\midrule
				
				\multicolumn{9}{l}{\textbf{Edge cut ratio (\%)}}\\
				\midrule
				
				\rowcolor{shade1}
				\textsc{HeiStream}
				& 50.74 & 52.57
				& 3.07 & 6.29
				& 42.85 & 42.15
				& 1.62 & 11.16 \\
				
				\rowcolor{shade2}
				\textsc{Cuttana}
				& 50.99 & 43.24
				& 22.31 & 1.37
				& 34.04 & 34.59
				& 21.70 & 3.24 \\
				
				\rowcolor{shade1}
				\textsc{BuffCut}
				& 49.32 & 44.79
				& 1.45 & 1.30
				& 27.95 & 27.92
				& 1.33 & 2.74 \\
				
				\specialrule{1.2pt}{2pt}{2pt}
				
				\rowcolor{shade2}
				\textsc{HeiStream-Re}
				& 47.98 & 50.67
				& \cellcolor{best}\textbf{0.34} & \cellcolor{best}\textbf{0.72}
				& 36.36 & 36.29
				& \cellcolor{best}\textbf{0.84} & 2.34 \\
				
				\rowcolor{shade1}
				\textsc{BuffCut-Re}
				& \cellcolor{best}\textbf{46.23} & \cellcolor{best}\textbf{41.92}
				& 1.27 & 1.04
				& \cellcolor{best}\textbf{25.31} & \cellcolor{best}\textbf{26.08}
				& 1.07 & \cellcolor{best}\textbf{1.26} \\
				
				\midrule
				\multicolumn{9}{l}{\textbf{Running time (s)}}\\
				\midrule
				
				\rowcolor{shade1}
				\textsc{HeiStream}
				& \cellcolor{best}\textbf{157.84} & \cellcolor{best}\textbf{136.74}
				& \cellcolor{best}\textbf{477.45} & \cellcolor{best}\textbf{365.39}
				& \cellcolor{best}\textbf{14.95}  & \cellcolor{best}\textbf{12.98}
				& \cellcolor{best}\textbf{44.17}  & \cellcolor{best}\textbf{36.60} \\
				
				\rowcolor{shade2}
				\textsc{Cuttana}
				& 670.71 & 507.60
				& 898.98 & 767.74
				& 251.99 & 259.68
				& 253.59 & 256.31 \\
				
				\rowcolor{shade1}
				\textsc{BuffCut}
				& 317.49 & 340.92
				& 741.63 & 550.30
				& 46.15  & 45.96
				& 62.03  & 55.93 \\
				
				\specialrule{1.2pt}{2pt}{2pt}
				
				\rowcolor{shade2}
				\textsc{HeiStream-Re}
				& 407.55 & 376.63
				& 1019.65 & 846.11
				& 39.77 & 35.60
				& 90.47 & 80.20 \\
				
				\rowcolor{shade1}
				\textsc{BuffCut-Re}
				& 517.06 & 504.45
				& 1310.19 & 1060.54
				& 68.67 & 67.50
				& 120.23 & 98.09 \\
				
				\bottomrule
			\end{tabular}
		}
	\end{table}
	
	\begin{figure}[t]
		\centering
		\includegraphics[width=1\linewidth]{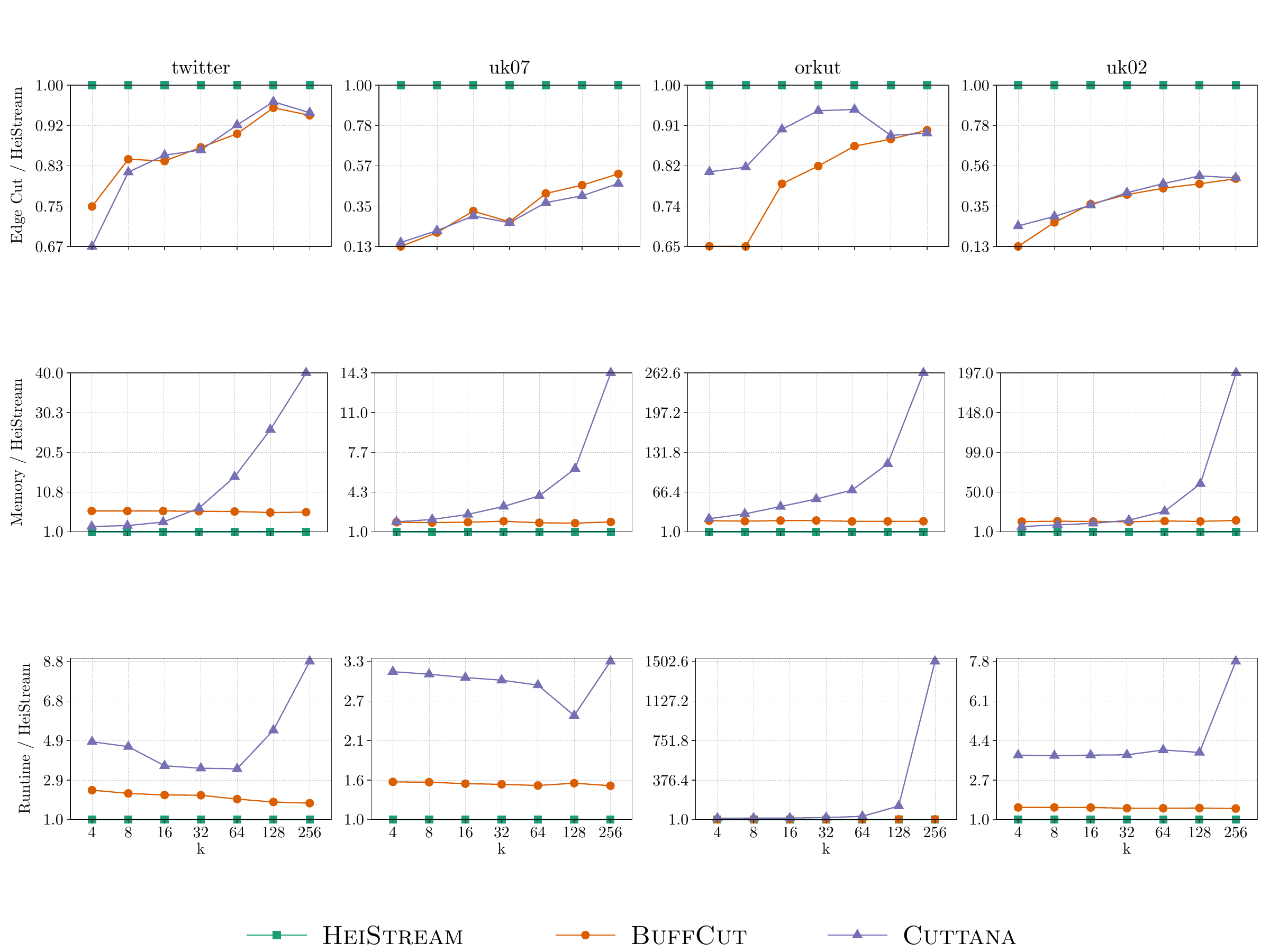}
		\caption{\textbf{Baseline comparisons under KONECT orderings (matching~\cite{cuttana}).} Normalized edge cut, runtime, and peak memory versus $k$ relative to \textsc{HeiStream}.}
		\label{fig:cuttana_baseline_graphs}
	\end{figure}
	\subparagraph*{KONECT Ordering.} 
	To match Hajidehi et al.'s~\cite{cuttana} experimental evaluation, we present results on the graphs found in the KONECT repository~\cite{konect}, marked with \textsuperscript{\dag} in Table~\ref{tab:datasets}, on both their source ordering and KONECT ordering. We run \ouralg with \hbox{$\mathcal{Q}_{\max}=2{,}097{,}152$} and~\hbox{$\delta=262{,}144$}. 
	We configure \textsc{Cuttana} and \textsc{HeiStream} with the parameters set by Hajidehi et al.~\cite{cuttana} and use an imbalance of $\varepsilon=5\%$ to match their experimental setup.
	Although~\cite{cuttana} reports \textsc{Cuttana} to be more efficient than \textsc{HeiStream}, our results show that \textsc{Cuttana}’s gains in quality come at a substantial computational cost. On the KONECT orderings, \textsc{Cuttana} improves the edge cut by between 17.7\% and 78.2\% relative to \textsc{HeiStream}, but is between 2.1$\times$ and 20.0$\times$ slower (Table~\ref{tab:k8_quality_runtime}). 
	Moreover, \textsc{Cuttana} exhibits significantly higher memory consumption as $k$ increases, using, for example, up to 260$\times$ more memory than \textsc{HeiStream} at $k=256$ (Figure~\ref{fig:cuttana_baseline_graphs}).
	\textsc{HeiStream}'s weaker performance in these experiments is largely attributable to stream ordering rather than inherent graph structure. The KONECT repository induces a first-appearance renumbering of node IDs that substantially reduces stream locality. When evaluated on the same graphs in their source orderings, \textsc{HeiStream} outperforms \textsc{Cuttana} in three out of four instances.
	In contrast, \ouralg is explicitly designed to be robust to adverse stream orderings, outperforming both \textsc{Cuttana} and \textsc{HeiStream} on all instances, except \texttt{twitter} under KONECT ordering. On KONECT ordering, it improves upon \textsc{Cuttana}'s solution quality by up to 20.0\% while using substantially fewer resources. \ouralg is faster across all four instances, with memory consumption remaining effectively constant as $k$ increases, while \textsc{Cuttana}'s resource consumption escalates sharply for $k > 32$ (Figure~\ref{fig:cuttana_baseline_graphs}). 
	With one additional streaming pass, \ouralg achieves lower edge cut than \textsc{Cuttana} across all KONECT instances and is faster in three of four instances.

	\begin{takeaway}
		\textbf{Takeaways.} \ouralg achieves the best solution quality, 15.8\% and 20.8\% better than \textsc{HeiStream} and \textsc{Cuttana} respectively, on the low-locality, randomly ordered Test Set, while being 2.9$\times$ faster and using 11.3$\times$ less memory than \textsc{Cuttana}, with modest overheads of 1.8$\times$ runtime and 1.09$\times$ memory over \textsc{HeiStream}. On KONECT-ordered instances, \ouralg achieves comparable or better edge cut (up to 20.0\%), while being faster on all instances and significantly more memory-efficient with increasing $k$.
		
	\end{takeaway}
	
	\section{Conclusion}
	
	We present \ouralg, a buffered streaming graph partitioner for computing high-quality balanced $k$-way partitions under strict memory constraints on adversarial stream ordering. Our approach uses prioritized batching: a bounded priority buffer actively regulates the order in which nodes are committed, delaying poorly informed decisions and incrementally forming batches with high internal locality, which are then partitioned using a multilevel scheme. Additionally, we introduce a refined buffer score that balances degree and available neighborhood information, and present a parallel implementation of \ouralg that achieves~1.8$\times$ faster runtime. Experiments on real-world and synthetic graphs under random node orderings demonstrate that,  in geometric means, \ouralg achieves 20.8\% fewer edge cuts while being 2.9$\times$ faster and using 11.3$\times$ less memory than the state-of-the-art prioritized buffering baseline, \textsc{Cuttana}. It produces 15.8\% better solution quality than the next-best buffered streaming partitioner, \textsc{HeiStream}, with modest overheads of 1.8$\times$ runtime and 1.09$\times$ memory. Compared to \textsc{Cuttana} on the KONECT-ordered instances evaluated by its authors, \ouralg is faster, uses less memory with increasing $k$, and achieves up to~20\% improvement in solution quality. In future work, we aim to extend prioritized buffered streaming to dynamic graphs and distributed-memory settings. Our implementation of \ouralg is integrated into \textsc{HeiStream} and is available as open source at \url{https://github.com/KaHIP/HeiStream}.
	
	\vfill
	
	\bibliography{compactfixed}

\end{document}